\documentclass[aps,prb,twocolumn,groupedaddress,bibnotes,citeautoscript]{revtex4-1}

\usepackage{graphicx}
\usepackage{dcolumn}
\usepackage{bm}
\usepackage{color}

\begin{document}


\title{Transport properties of KTaO$_3$ from first-principles}

\author{Burak Himmetoglu}
\affiliation{Enterprise Technology Services, University of California, Santa Barbara, CA 93106}
\author{Anderson Janotti}
\affiliation{Materials Science and Engineering, University of Delaware, Newark, DE 19716}

\date{\today}

\begin{abstract}

The transport properties of the perovskites KTaO$_3$ are calculated using first-principles methods. 
Our study is based on Boltzmann transport theory and the relaxation time approximation,
where the scattering rate is calculated using an analytical model describing 
the interactions of electrons and longitudinal optical phonons. We compute the 
room-temperature electron mobility and Seebeck coefficients of KTaO$_3$, and SrTiO$_3$ for comparison, 
for a range of electron concentrations.
The comparison between the two materials provides insight into the mechanisms that determine
room-temperature electron mobility, such as the effect of band-width and spin-orbit 
splitting. The results, combined  with the efficiency of the computational scheme developed in this study, provide 
a path to investigate and discover materials with targeted transport properties.
\end{abstract}


\maketitle

\section{Introduction}

KTaO$_3$ (KTO) is a member of a class of materials with the perovskite crystal structure which are being explored 
in the emerging area of oxide electronics. 
These materials display an array of interesting physical phenomena, that includes ferroelectricity, magnetism, metal-insulator transition 
and superconductivity~\cite{hwang-emergent}.
Such remarkable properties are largely related to the strongly localized $d$ orbitals that dominate the electronic structure of these
materials.  An immediate drawback for electronic applications is the limitation in electron mobility, 
which is often attributed to the heavy band masses resulting from $d$-orbital derived
conduction bands and enhanced electron-phonon interactions.  
Low-temperature electron mobilities exceeding 20,000 cm$^2$/Vs in KTO have been reported, 
while the observed room-temperature mobilities are limited to 30 cm$^2$/Vs~\cite{kto-mobility}.
In the related and more widely studied material SrTiO$_3$ (STO), low-temperature mobilities exceeding 50,000 cm$^2$/Vs 
have been reported, while mobilities at room temperature
are less than 10 cm$^2$/Vs~\cite{stemmer-new,stemmer-sto-mobility}.

The realization of rather high density two-dimensional electron gas at the interface
of perovskite materials~\cite{ohtomo-2deg,moetakef-2deg} has led to an increased
interest in these systems for device applications. However, limits in room-temperature 
electron mobility constitute a major hurdle. On the other hand, these materials
display notably large Seebeck coefficients~\cite{sto-seebeck-kuomoto,sto-seebeck,kto-S,kto-mobility,stemmer-seebck}, 
which makes them promising for thermoelectric applications.  
In this context, a first-principles approach to the transport properties in these materials is crucial
to uncover the fundamental microscopic mechanisms behind the experimental observations,
and pave the way to improved device performance.

At very low temperatures (below 10 K), the electron mobility is limited by ionized 
impurity scattering~\cite{Jena-STO,Spinelli-STO}.  At intermediate to low temperatures, 
recent studies have indicated that the mobility is determined by a combination of several mechanisms, such as 
transverse optical (TO) phonon scattering~\cite{Jena-STO} and electron-electron scattering~\cite{stemmer-sto-T2}.
However, the rather low mobility at room temperature is generally believed to originate from 
strong scattering of electrons by polar longitudinal optical (LO) phonons~\cite{frederikse-sto,tufte-sto}. 
In general, studies of electron mobility are based on experimental analysis of transport data using phenomenological models,
whereas investigations using first-principles methods have been scarce. 

Recent theoretical studies indicate that LO phonons are the main source of 
scattering which leads to the observed low mobilities at room temperature~\cite{sto-us,polaron-devreese}.
Investigations of other mechanisms, such as TO phonon scattering, have been limited 
due to difficulties arising from strong anharmonic effects at the temperature range (around 100K) where
they become effective~\cite{sto-ph3,siemons-kto}. There are some
limited studies on the effect of electron-electron scattering in STO~\cite{Mazin-T2,sto-ee},
however, a study based on a fully first-principles band structure is still not yet available.
In the case of thermal transport, few groups have investigated 
the Seebeck coefficient of STO and KTO~\cite{seebeck-sto,seebeck-stokto}, 
using calculated effective masses and density of states.
Although the results are in reasonable agreement with experiments, the effect of electron-phonon scattering
on the thermal transport properties has not been explicitly included in these studies.

In this work, we investigate the transport properties of KTO and STO, and 
analyze the underlying mechanisms that determine the room-temperature
transport coefficients. Our calculations are based on a fully first-principles
description of the electronic band structure of both materials, and a solution
of the Boltzmann transport equation (BTE) within the relaxation time approximation.
The relaxation time, i.e., inverse scattering rate, is calculated by an analytical model for the electron-(LO)phonon scattering,
providing an efficient computational approach compared to calculations of electron-phonon 
scattering matrix elements using density functional perturbation theory~\cite{restrepo,kaasbjerg-mos2,mbn-1,mbn-2,wuli}. 
The comparison of the results for STO and KTO show that the band-widths of the conduction bands as well as the spin-orbit splitting play
an important role in determining the room-temperature transport properties.
%

Our results show that the superior room-temperature mobility in KTO, compared to STO, arises not only from 
the larger conduction band-width, but also from the stronger spin-orbit coupling which lifts the degeneracy of the lowest-energy conduction bands, 
thus reducing the effective number of conduction bands to which electrons can scatter. 
Therefore, the lower effective number of conduction bands mitigates the strength of electron-phonon scattering, leading 
to a significant enhancement of carrier mobility at room temperature. On the other hand, it also
yields a smaller Seebeck coefficient.  Such insights, combined with the efficiency of computational approach described here, opens up the possibility to
screen a large number of perovskite oxides aiming at the discovery of new materials with targeted transport properties. 

The paper is organized as follows: In Section~\ref{sec:methods}, we describe the details of the computational approach. 
In Section~\ref{sec:bandsph}, we report the calculated electronic band structure and
optical phonon frequencies of both KTO and STO. In Section~\ref{sec:eph}, we discuss details on the calculation 
of electron-phonon scattering rates, and in Section~\ref{sec:tr}, we present the calculations of transport 
integrals and discuss the behavior of mobility and Seebeck coefficient as a 
function of electron concentration in KTO and STO. In Section~\ref{sec:tauk}, we provide 
a detailed discussion of the differences in transport coefficients arising from 
using a constant scattering time {\em vs.} the full ${\bf k}$-dependent scattering time obtained
from electron-phonon scattering rates. In Section~\ref{sec:disc} we present our
concluding remarks.

\section{\label{sec:methods} Computational Methods}

Structural optimizations and electronic band structure calculations presented in this
paper are performed using density functional theory (DFT) and the plane-waves pseudopotential method
as implemented in the PWSCF code of the {\it Quantum ESPRESSO} package~\cite{QE}.
The exchange-correlation energy is approximated using the local density approximation (LDA)
with the Perdew-Zunger parametrization~\cite{pz}.  K, Sr, Ta, Ti and O atoms are represented by 
ultrasoft pseudopotentials~\cite{uspp}.  Fully relativistic pseudopotentials are employed with non-collinear 
spins~\cite{dalcorso} in the calculations including spin-orbit coupling.
The electronic wavefunctions and charge density are
expanded in plane-wave basis sets with energy cutoffs of 50 Ry and 600 Ry, respectively. 
Brillouin-zone (BZ) integrations are performed on a 8$\times$8$\times$8 grid of special $k$-points\cite{mp}. 
The phonon frequencies at the zone center ($\Gamma$) are calculated using 
density functional perturbation theory (DFPT) as implemented in {\it Quantum ESPRESSO}~\cite{dfpt}.
The splitting of longitudinal and transverse optical modes at $\Gamma$ is taken into account 
via the method of Born and Huang~\cite{born-huang}.
For the calculation of the electron-phonon scattering rates and transport integrals, a 60$\times$60$\times$60
grid is used to sample the BZ. The Dirac-delta functions appearing in the calculation of scattering
rates is approximated by a Gaussian smearing function with a width of 0.2 eV. 

\section{\label{sec:bandsph}Electronic and Vibrational Spectrum}

We consider the cubic phase of KTaO$_3$, which is the stable phase at room temperature. The optimized lattice parameter
is $a_0 = 3.94$ \AA, which is about $1\%$ smaller than the experimental value~\cite{kto-lattice}, as expected from the LDA functional.
LDA calculations yield an indirect band gap of 2.13 eV for the collinear spin, and
2.00 eV for the non-collinear spin calculations, respectively. As expected,
LDA underestimates the band gap of KTO, which is 3.64 eV (indirect)~\cite{kto-gap}. 
We note that the band gap does not enter explicitly in the transport calculations, 
and only the conduction band structure is relevant, as discussed previously in Ref.(\onlinecite{sto-us}).
Therefore, we limit our study to the LDA results in this work.

\begin{figure}[!ht]
\includegraphics[width=0.35\textwidth]{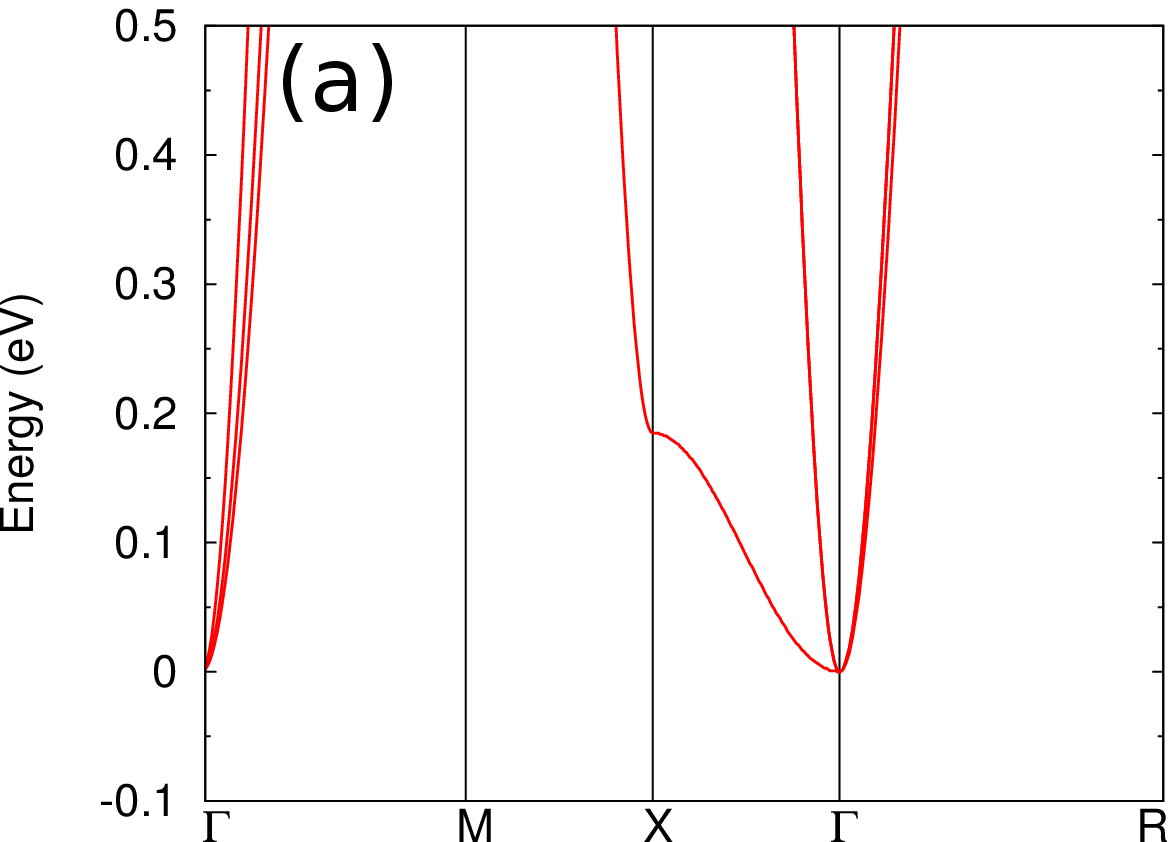}
\includegraphics[width=0.35\textwidth]{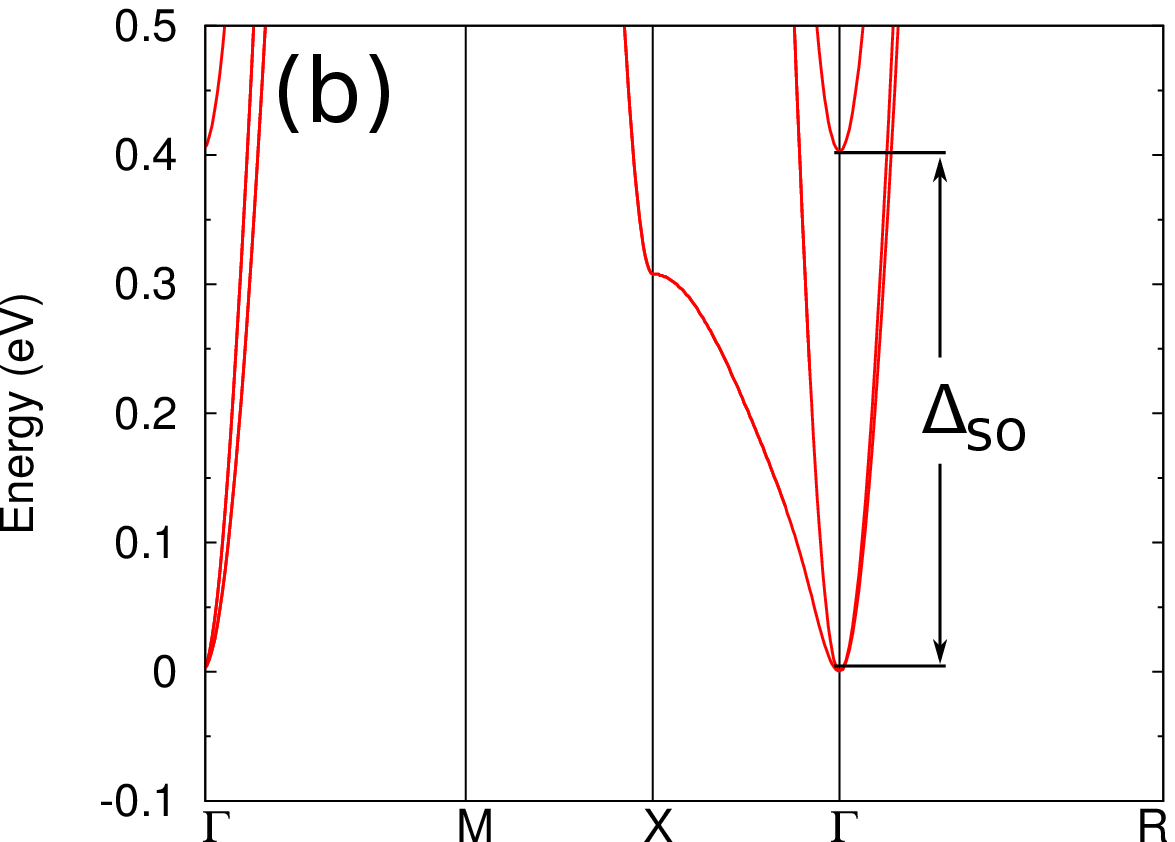}
\hspace{0.5cm}
\includegraphics[width=0.35\textwidth]{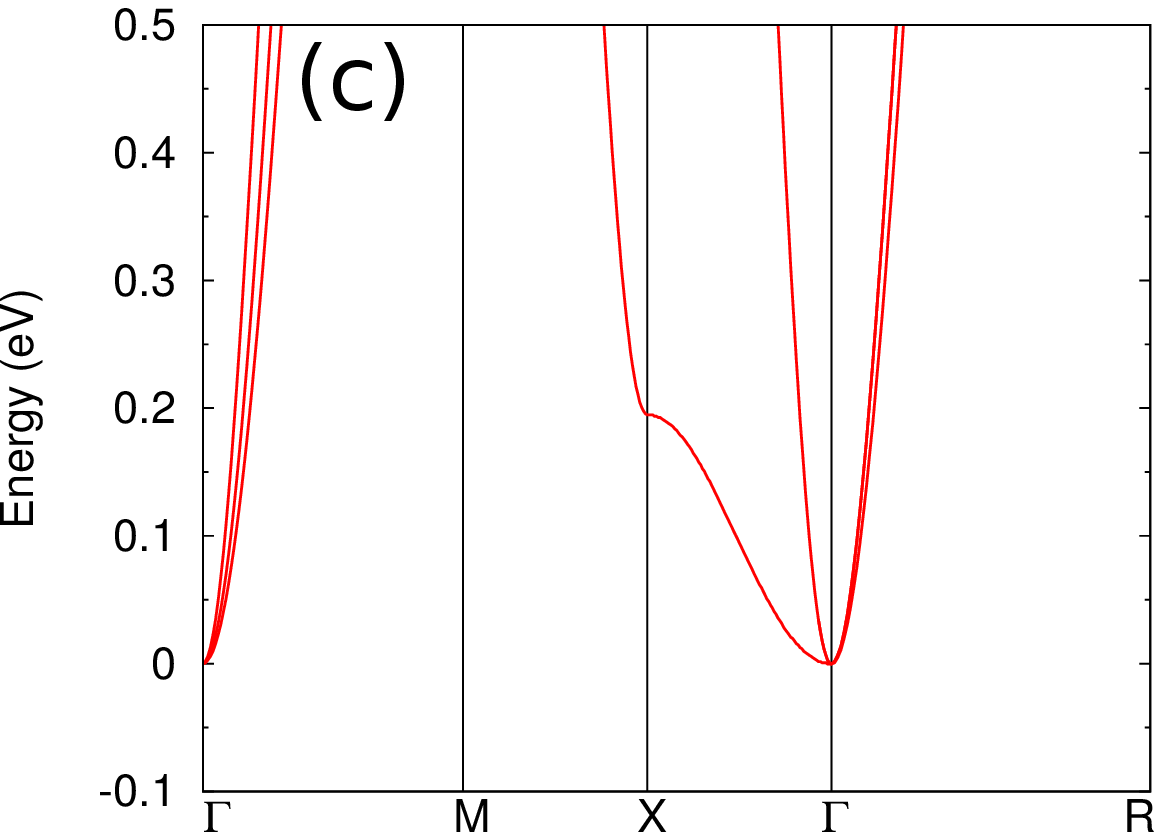}
\caption{\label{fig:bands} Conduction band structure for KTO, (a) in collinear calculation,
and (b) in  noncollinear calculation. Band structure of STO is shown in (c) for comparison.}
\end{figure}

The conduction band structure of KTO is shown in Fig.~\ref{fig:bands}, where
results are presented for both collinear and non-collinear calculations. 
Also shown is the band structure of STO (from Ref.(\onlinecite{sto-us})) for comparison.
The conduction bands are derived mainly from the Ta 5$d$ orbitals. Due to cubic symmetry,
the 5$d$ states are split into e$_g$ and t$_{2g}$ states, with the lowest lying conduction bands being t$_{2g}$ states.
In case of collinear spin calculation, spin-orbit interactions are not present,
and therefore the three conduction bands are degenerate at $\Gamma$. In non-collinear spin
calculations, spin-orbit coupling is included, the threefold degeneracy at $\Gamma$ is lifted,
and one of the bands is split by $\Delta_{\rm SO} \simeq 0.4 {\rm eV}$ to higher energy (see Fig.~\ref{fig:bands}(b)).  

The phonon frequencies at $\Gamma$ and the electronic dielectric
constant were obtained by DFPT calculations. The calculated electronic dielectric constant 
is $\epsilon_{\infty} = $ 5.4, which is in reasonable agreement with the
experimental value of approximately 4.6~\cite{kto-dielectric}. 
The calculated phonon frequencies at $\Gamma$ are listed in Table~\ref{tab:optph}, 
which are also in good agreement with the experimental measurements and
with previous calculations~\cite{kto-singh}.

For STO, we use the phonon frequencies and the electronic structure reported in Ref.~\onlinecite{sto-us}, 
which were also calculated using LDA. 
Unlike KTO, there is negligible difference between collinear and non-collinear calculations in STO, 
due to Ti having a much smaller atomic mass than Ta, which leads to
a relatively small spin-orbit splitting of 28 meV~\cite{sto-hse-bs}.
Therefore, only collinear calculations for STO are used as a basis of comparison. 


\begin{table}[!ht]
\caption{\label{tab:optph} Calculated and experimental longitudinal optical (LO)
and transverse optical (TO) phonon frequencies at the $\Gamma$ point,
in units of cm$^{-1}$.}
\begin{ruledtabular}
\begin{tabular}{cccc}
   \multicolumn{2}{c}{ LO} & \multicolumn{2}{c}{TO} \\
  LDA & Exp. & LDA & Exp. \\\hline
 176 & 188$^a$, 184-185$^{b,c}$ & 105  &  88$^a$, 81-85$^b$, 88$^c$ \\
 253 & 290$^a$, 279$^b$, 255$^c$ & 186 & 199$^a$, 198-199$^{b,c}$ \\
 403 & 423$^a$, 421-422$^{b,c}$ & 253 & 290$^a$, 279$^b$, 255$^c$ \\
 820 & 833$^a$, 826-838$^{b,c}$ & 561 & 549$^a$, 546-556$^b$, 574$^c$ 
\end{tabular}
\end{ruledtabular}
\begin{flushleft}
$^a$ Ref.~\onlinecite{kto-phon-1}, 
$^b$ Ref.~\onlinecite{kto-phon-2}, 
$^c$ Ref.~\onlinecite{kto-phon-3}.
\end{flushleft}\end{table}

\section{\label{sec:eph}Electron-phonon coupling}

In this work, we consider electron-phonon interactions only due to polar longitudinal optical (LO) phonons.
The interactions of electrons with the macroscopic electric field due to a LO mode can be analytically 
described by the Fr\"ohlich model~\cite{mahan}, where the electron-phonon coupling depends on 
the dielectric constants and the LO phonon frequency. In STO and KTO, there are multiple LO modes
that contribute to the macroscopic electric field.  
In this case, the electron-phonon coupling matrix element squared is given by~\cite{polaron-devreese,toyozawa,eagles}
\begin{equation}
\left\vert g_{{\bf q}\, \nu} \right\vert^2 = \frac{1}{q^2}\, 
  \left( \frac{e^2\, \hbar\, \omega_{L\nu}}{2 \epsilon_0\, V_{\rm cell}\, \epsilon_{\infty}} \right)\,
  \frac{\prod_{j=1}^{N} \left(1-\frac{\omega_{T j}^2}{\omega_{L \nu}^2}\right)}
       {\prod_{j \neq \nu} \left( 1 - \frac{\omega_{L j}^2}{\omega_{L \nu}^2} \right)},
\label{eqn:gnu2}
\end{equation}
where N is the number of LO/TO modes, $\omega_{L \nu}$ are LO mode frequencies,
$\omega_{T \nu}$ are TO mode frequencies, V$_{\rm cell}$ is the unit cell volume,
and $\epsilon_{\infty}$ is the electronic dielectric constant. 
This equation is obtained using the electron-phonon Hamiltonian 
in Ref.~\onlinecite{toyozawa} and the Lyddane-Sachs-Teller (LST) relation
$\epsilon(\omega) = \epsilon_{\infty}\, \prod_{j=1}^{N}\, (\omega^2-\omega_{L,j}^2)/(\omega^2-\omega_{T,j}^2)$.
In this expression,
the dispersion of the optical phonon frequencies are ignored and in our calculations,
we use their value at the zone center. Alternative formulations of 
the LO phonon coupling based on Born effective charges, which take into account their full dispersion,
have also been discussed in the literature~\cite{calandra-ep,giustino-ep}.  
 The calculated electron-phonon coupling elements are listed in Table~\ref{tab:Cnu}.  Here, we define 
\begin{equation}
C_{\nu} = \left(\frac{a_0}{2\pi}\right)^2\, q^2\, \vert g_{{\bf q} \nu} \vert^2,  \label{eqn:Cnu}
\end{equation}
where $a_0$ is the cubic lattice parameter.
\begin{table}[!ht]
\caption{\label{tab:Cnu} Calculated electron-phonon couplings in $eV^2$ 
based on Eqns.(\ref{eqn:gnu2}) and (\ref{eqn:Cnu}).}
\begin{ruledtabular}
\begin{tabular}{ccc}
        & STO                   & KTO \\
\hline
  $C_1$ & $1.43 \times 10^{-5}$ & $1.82 \times 10^{-5}$ \\ 
  $C_2$ & $1.33 \times 10^{-3}$ & $1.46 \times 10^{-3}$ \\
  $C_3$ & $6.77 \times 10^{-3}$ & $7.50 \times 10^{-3}$ 
\end{tabular}
\end{ruledtabular}
\end{table}
Note that there are only three coupling constants listed in Table~\ref{tab:Cnu}.
One of the LO modes in Table~\ref{tab:optph} is not polar-- the mode with frequency 253 cm$^{-1}$
appears both in the LO and TO sectors, and since it is nonpolar, it 
does not split up. The constants $C_{\nu}$ are enumerated in order of increasing 
LO mode frequency. The electron-phonon couplings reported here for 
STO are lower than those reported in Ref~\onlinecite{sto-us}. This is due to the fact 
that the electron-phonon coupling constant used in Ref~\onlinecite{sto-us} was
only valid for the case of one LO mode, overestimating their strength 
when three modes were present.  

The coupling constants listed in Table~\ref{tab:Cnu} are also
in agreement with previous results in literature available for STO~\cite{polaron-devreese}. 
For comparison, we computed values of $C_{\nu}$ resulting from
the parameters reported in Ref~\onlinecite{polaron-devreese}. Using the 
dimensionless constants $\alpha_{\nu}$, experimental phonon frequencies,
and the band mass of $0.81\, m_e$, from Ref~\onlinecite{polaron-devreese},
one obtains:
$C_1 = 1.26 \times 10^{-5}\, eV^2$, $C_2 = 1.25 \times 10^{-3}\, eV^2$ and
$C_3 = 9.50 \times 10^{-3}\, eV^2$. 
The first two coupling constants are in good agreement with our 
results for STO in Table~\ref{tab:Cnu}. Our constant $C_3$
is slightly lower than that of  Ref~\onlinecite{polaron-devreese}. 
The effective mass
enters into the definition of the dimensionless constants $\alpha_{\nu}$
defined in Ref~\onlinecite{polaron-devreese}, 
but cancels out when $C_{\nu}$ are calculated, so its value does not have any effect.
The differences between our results and that of Ref~\onlinecite{polaron-devreese} come from the fact that we use calculated phonon
frequencies and dielectric constants instead of experimental values. 

Using Fermi's golden rule, one can write the scattering rate of electrons 
in the state $\psi_{n {\bf k}}$ as~\cite{mahan}
\begin{eqnarray}
&& \tau_{n {\bf k}}^{-1} = \frac{2\pi}{\hbar}\, \sum_{{\bf q} \nu, m}\, \vert g_{{\bf q} \nu} \vert^2\, \times
    \nonumber\\
&& \quad \Big\{ \left( n_{{\bf q} \nu} + f_{m, {\bf k+q}} \right)\, \delta\left( \epsilon_{m, {\bf k+q}} - \epsilon_{n {\bf k}} - \hbar \omega_{L \nu} \right)
  \nonumber\\
&& \quad + \left( 1 + n_{{\bf q} \nu} - f_{m, {\bf k+q}} \right)\, \delta\left( \epsilon_{m, {\bf k+q}} - \epsilon_{n {\bf k}} + \hbar \omega_{L \nu} \right)
\Big\}, \label{eqn:tau}
\end{eqnarray}
where $\epsilon_{n {\bf k}}$ are electron energies, $\hbar \omega_{L \nu}$ are LO phonon energies,
$n_{{\bf q} \nu}$ and $f_{m, {\bf k+q}}$ are phonon and electron occupation factors described by Bose-Einstein and Fermi-Dirac
distributions, respectively. More precisely, the rate in Eqn. (\ref{eqn:tau}) corresponds to the quasiparticle relaxation rate for electrons
interacting with LO phonons, and the transport rate contains extra band-velocity factors~\cite{mahan}. However, recent studies
 have shown that these factors do not lead to significant changes in the calculated transport properties~\cite{wuli,sto-us},
therefore we use the form in Eqn.(\ref{eqn:tau}) for brevity.

\begin{figure}[!ht]
\includegraphics[width=0.45\textwidth]{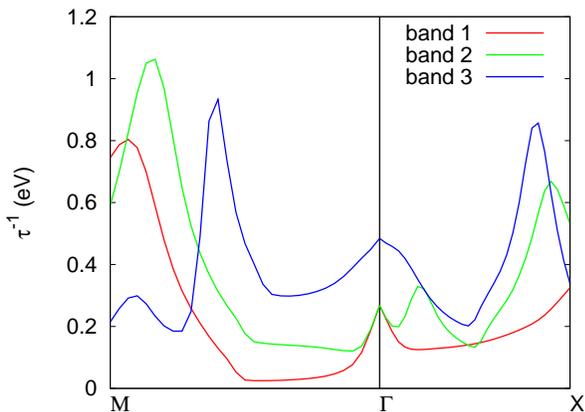}
\caption{\label{fig:tauBZ} Electron-phonon scattering rate $\tau^{-1}_{n {\bf k}}$ calculated 
along $\Gamma$-X and $\Gamma$-M for the three conduction bands at 300 K and $n=10^{20} \rm{cm}^{-3}$.}
\end{figure}

In Fig.~\ref{fig:tauBZ} we show the scattering rates $\tau_{n {\bf k}}^{-1}$ along $\Gamma$-X and $\Gamma$-M. 
The rates for the three t$_{2g}$-derived conduction bands in KTO are shown.
Each of the three conduction bands depicted represents a doubly degenerate state. 
For all the conduction bands, the rate initially decreases away from $\Gamma$, and then 
raises again towards the middle of the zone. 
%

The dependence of the scattering rates on electron density can be obtained by defining the following quantity
\begin{equation}
D_n\, \tau_n^{-1}(E_F) \equiv \sum_{\bf k}\, \delta( E_f - \epsilon_{n {\bf k}} )\, \tau_{n {\bf k}}^{-1}
\label{eqn:Dtau}
\end{equation}
where $D_n(E_F) = \sum_{\bf k}  \delta( E_f - \epsilon_{n {\bf k}} )$ is the density of states
at the Fermi level. $D_n \tau^{-1}_n(E_F)$ represent the scattering rate for each band weighted 
by the density of states (it is dimensionless when $\tau^{-1}$ is represented in units of energy). 
Since the scattering rate near Fermi level 
determines the transport properties, (see Eqns. (\ref{eqn:int})), $D_n \tau^{-1}_n(E_F)$ represents a measure 
of the strength of electron-phonon scattering for each band. 

\begin{figure}[!ht]
\includegraphics[width=0.4\textwidth]{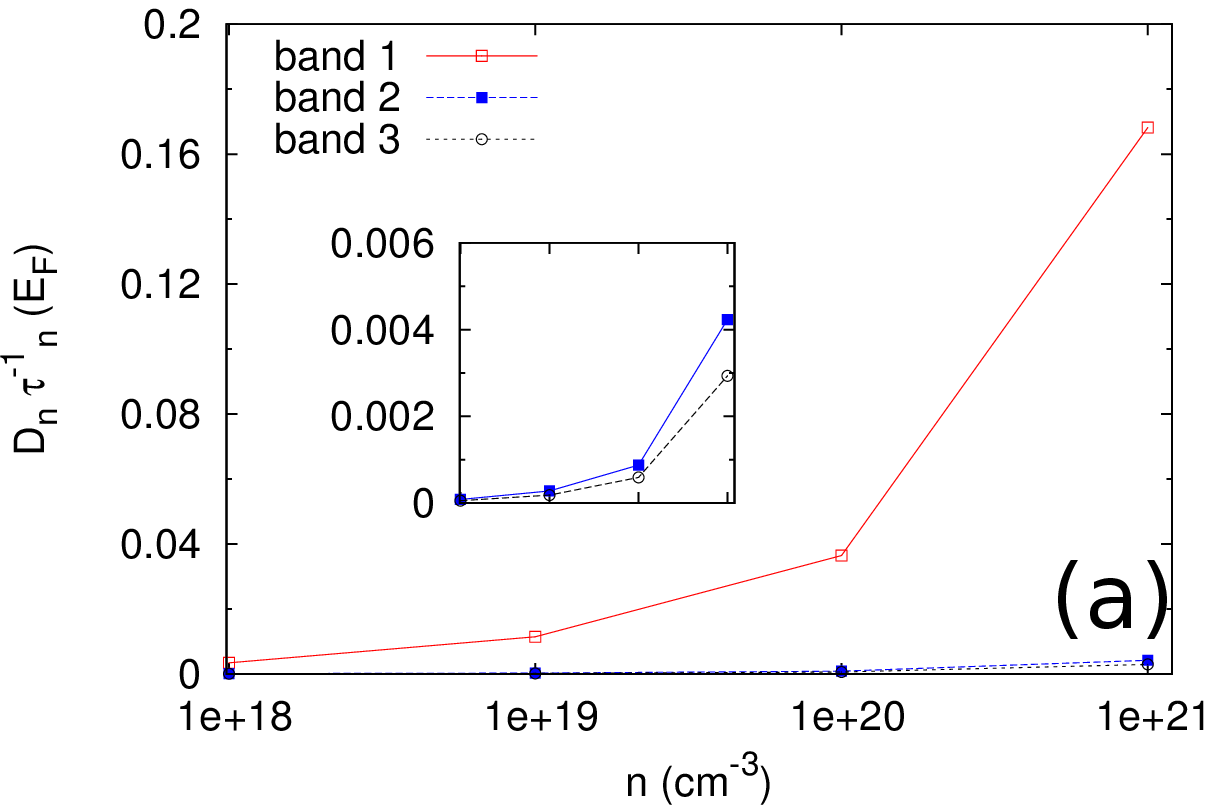}
\includegraphics[width=0.4\textwidth]{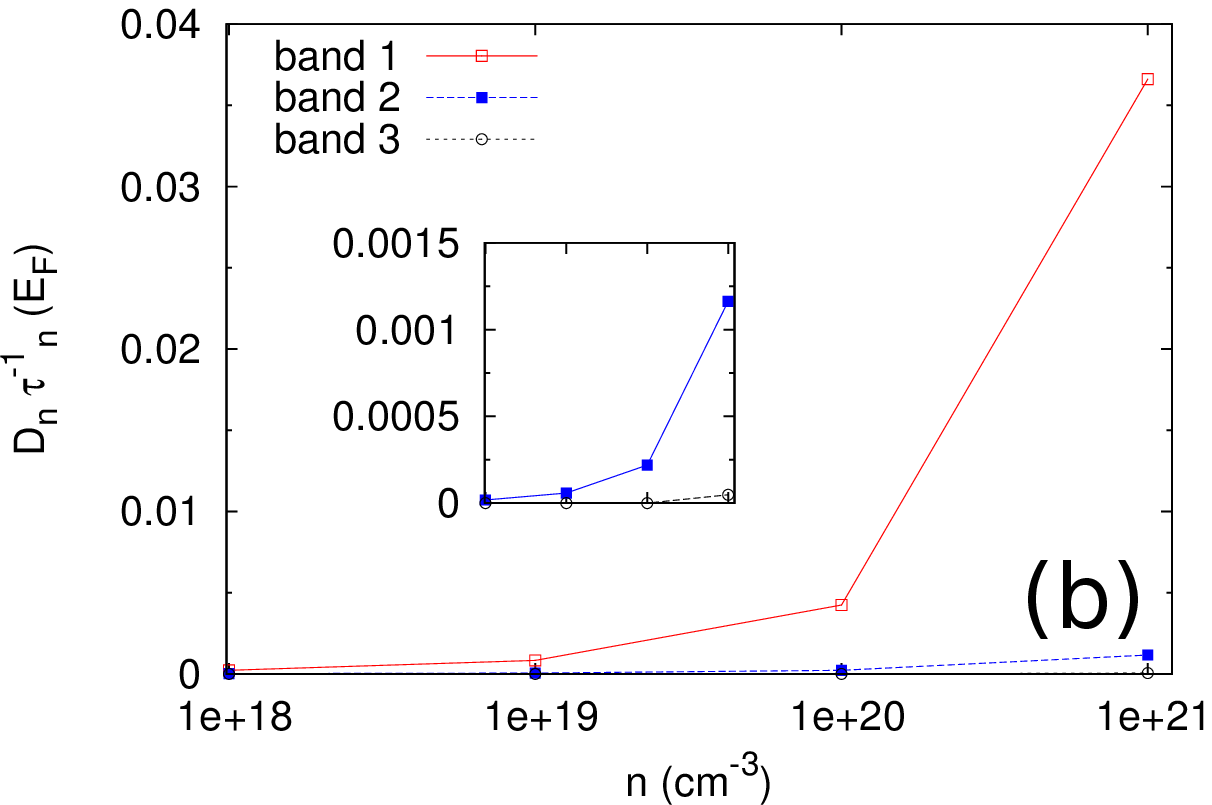}
\caption{\label{fig:tauEf} Electron-phonon scattering rates in KTO weighted by the density of states
at the Fermi level for (a) collinear and (b) noncollinear calculation as a function of
electron density at 300 K.}
\end{figure}

\begin{figure}[!ht]
\includegraphics[width=0.45\textwidth]{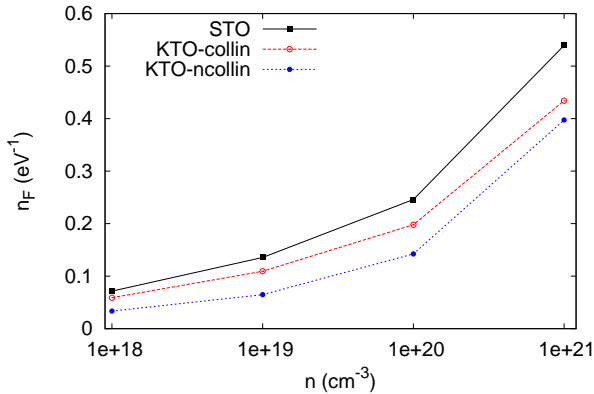}
\caption{\label{fig:nef} Densities of states at the Fermi level. For KTO,
both collinear and noncollinear calculations are displayed.}
\end{figure}

The weighted scattering rates at the Fermi level for each of the conduction bands are shown in Fig.~\ref{fig:tauEf}.
The lowest lying band (i.e. band 1) always has the highest rate, as a result of the enhancement 
in the density of states, since this band is the less dispersive among the three. Comparing the collinear and noncollinear
calculations, we observe that inclusion of spin-orbit coupling reduces the effective rate at E$_F$ significantly. 
This is due to two factors: 
i) Since the highest lying band is pushed up in energy by 0.4 eV, electrons residing in the two lower lying 
bands cannot scatter via LO phonons to this band (and vice versa), reducing the available scattering channels. The spin-orbit 
splitting is much larger than the maximum LO phonon energy (around 0.1 eV), and as a result the energy conserving delta
functions in Eq.(\ref{eqn:tau}) vanish for inter-band transitions involving the split-off band. 
ii) Collinear spin calculations result in less dispersive bands resulting in higher density of states,
as can be seen in Fig.~\ref{fig:nef}.
This effect has also been observed for effective band masses near the zone center~\cite{sto-hse-bs}. 

The inset in Fig.~\ref{fig:tauEf}(b) shows that the split-off band (band 3) has significantly smaller effective scattering
rate at E$_F$ compared to the collinear case in Fig.~\ref{fig:tauEf}(a). This is due to the fact that the Fermi energies for the
considered electron densities are much lower than the split-off band energies, leading to a reduction of its density of states. 
The overall reduction of scattering rates when spin-orbit coupling is strong has in fact been proposed before~\cite{sto-us} and
KTO serves as an example for this behavior.

\section{\label{sec:tr}Transport Integrals}

For calculation of transport coefficients, we compute the following two integrals:

\begin{eqnarray}
&& I^{(1)}_{\alpha\, \beta} = \frac{1}{V_{\rm cell}}\, \sum_{n, {\bf k}}\,
  \tau_{n {\bf k}}\, \left( - \frac{\partial f_{n {\bf k}}}{\partial \epsilon_{n {\bf k}}} \right)\,
  v_{n {\bf k}, \alpha}\, v_{n {\bf k}, \beta} \nonumber\\
&& I^{(2)}_{\alpha\, \beta} = \frac{1}{V_{\rm cell}}\, \sum_{n, {\bf k}}\,
  \tau_{n {\bf k}}\, \left( - \frac{\partial f_{n {\bf k}}}{\partial \epsilon_{n {\bf k}}} \right)\,
  \left( \epsilon_{n {\bf k}} - E_F \right) \, v_{n {\bf k}, \alpha}\, v_{n {\bf k}, \beta}. \nonumber\\
\label{eqn:int}
\end{eqnarray}
In terms of these integrals, the conductivity ($\sigma$) and Seebeck ($S$) tensors are given by
\begin{equation}
\sigma_{\alpha \beta} = 2\, e^2\, I^{(1)}_{\alpha \beta} \,\,\, , \,\,\,
S_{\alpha \beta} = -\frac{1}{e T}\, I^{(1) -1}_{\alpha \nu}\, I^{(2)}_{\nu \beta}, \label{eqn:tr}
\end{equation}
where E$_F$ is the Fermi energy, $\alpha, \beta, \nu$ are cartesian indices (repeated indices are summed over), and
the factor 2 accounts for the spin (in case of the noncollinear calculation, the factor 2 is not included
and the band indices $n$ also contain the total angular momentum quantum number $j=l \pm 1/2$). 
The band velocities $v_{n {\bf k}}$ are defined as
\begin{equation}
v_{n {\bf k}, \alpha} \equiv \frac{1}{\hbar}\, \frac{\partial \epsilon_{n {\bf k}}}{\partial k_{\alpha}}.
\label{eqn:v}
\end{equation}
In this study, we compute band velocities in a very dense grid (60$\times$60$\times$60) using finite differences. 
For the cases of KTO and STO, the conduction band structure is rather simple (i.e. no band crossings), therefore
this approach leads to accurate band velocities. In more complex band structures, it is necessary to use
interpolation techniques for accurate calculations~\cite{boltztrap,boltzwann}.
Since KTO and STO are cubic, the conductivity and the Seebeck tensors are
diagonal with equal entries. Therefore, we will refer to the scalar quantities of conductivity ($\sigma$) and Seebeck coefficient (S)
from now on. The mobility is defined through the ratio of the conductivity to the electron density as 
\begin{equation}
\mu = \frac{\sigma}{n e}. \label{eqn:mob}
\end{equation}
Using the scattering rates computed from Eqn.(\ref{eqn:tau}), based on the electron-phonon couplings
of Eqn.(\ref{eqn:gnu2}), we calculate the transport integrals in Eqn.(\ref{eqn:int}) for KTO and STO. For KTO,
we perform the integration both for collinear and noncollinear band structures while for STO we only use the
collinear band structure.  We have explicitly checked that the noncollinear calculations in STO lead to
only negligible changes in the transport integrals. In the following two subsections, we report the the room-temperature 
mobility and Seebeck coefficients.
\begin{figure}[!ht]
\includegraphics[width=0.45\textwidth]{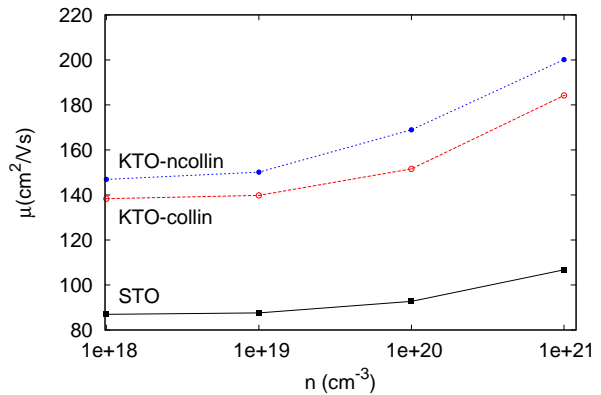}
\caption{\label{fig:muall} Calculated mobilities of STO and KTO at 300K. For KTO,
both collinear and noncollinear calculations are displayed. } 
\end{figure}

\subsection{Electron mobilities}

Fig.~\ref{fig:muall} shows the calculated mobilities as a function of electron concentration at 300 K. 
KTO with spin-orbit interactions taken into account (i.e. noncollinear) displays the highest mobility
across the range of electron densities we have considered, whereas the mobilities in STO are significantly lower
~\footnote{ In a previous work~\cite{sto-us}, the mobilities calculated at room temperature for 
STO are much smaller than the ones reported here. There are two reasons for this difference: 1)
In Ref.~\onlinecite{sto-us}, scattering rates are approximated with their values at the zone center,
2) The electron-phonon coupling was described by a different formula, which was only 
valid only for the case of a single LO mode. Instead, the correct formula is used in this paper.}.
As discussed previously, inclusion of spin-orbit splitting reduces the effective scattering rate, 
leading to an increase in the mobilities, which is the reason for higher values in the noncollinear calculation.
In the case of the collinear calculations, where the bands are not split due to spin-orbit coupling, 
the mobilities are higher in KTO than in STO because the higher band velocities.
Since Ta 5$d$ orbitals lead to wider conduction bands compared to
those of Ti 3$d$, such an increase is expected. These findings are in agreement with the experimental
observation of higher room-temperature mobilities in KTO compared to STO~\cite{kto-mobility,kto-S}.

We note that our calculations overestimate the observed room temperature mobilities of both STO, which is less than 10 cm$^2$/Vs 
~\cite{frederikse-sto,tufte-sto,stemmer-sto-mobility}, and KTO, which is around 30 cm$^2$/Vs~\cite{kto-mobility},
for a range of electron concentrations.
This is due to the fact that our calculations contain only LO phonon scattering mechanism, 
and ignores others such as impurities~\cite{Spinelli-STO}, TO-phonon~\cite{Jena-STO} and electron-electron scattering~\cite{stemmer-sto-T2}.

\begin{figure}[!ht]
\includegraphics[width=0.45\textwidth]{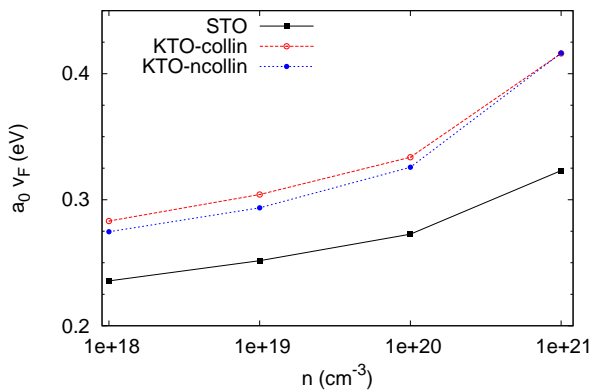}
\caption{\label{fig:v2f} Averaged Fermi velocities (rescaled with the lattice parameter $a_0$). For KTO,
both collinear and noncollinear calculations are displayed. }
\end{figure}

Further insight into the behavior of mobility shown in Fig.(\ref{fig:muall}) can be gained
by calculating the average values of the Fermi velocities, which we define as
\begin{equation}
v^2_F \equiv \frac{\sum_{n, {\bf k}} \delta (E_F - \epsilon_{n {\bf k}})\, v_{n {\bf k}}^2 }
                     {\sum_{n, {\bf k}} \delta (E_F - \epsilon_{n {\bf k}})},
\label{eqn:v2e}
\end{equation}
where the index $n$ runs over the conduction bands.
Shown in Fig.~\ref{fig:v2f} are $v_F$, which have higher values in KTO compared to STO.
While KTO in the noncollinear calculation has a higher band-width (as can seen in Figs.~\ref{fig:bands} and ~\ref{fig:nef}),
the averaged values of the Fermi velocities are slightly lower than that of KTO in the collinear calculation.
This is most probably due to the highly non-parabolic Fermi surface of KTO, which has high and low 
velocity regions, averaging out to a similar numerical value for both collinear and non-collinear calculations. 
Instead, lower scattering rates in the noncollinear calculations lead to the 
enhancement of mobilities as seen in Fig.~\ref{fig:muall}. 

An additional feature which deserves attention is the increase in mobility with increasing electron density $n$
as seen in Fig.~\ref{fig:muall}. This may seem counterintuitive,
since the effective scattering rates at E$_F$ also increases with $n$ (see Fig.~\ref{fig:tauEf}). However, band velocities $v_{n {\bf k}}$
increase as a function of electron energy in the conduction band.  
The increase in $v_{n {\bf k}}^2$ dominates over the increase in $\tau^{-1}_{n {\bf k}}$
leading to an overall increase of the mobility as a function of $n$.

\subsection{Seebeck coefficients}

\begin{figure}[!ht]
\includegraphics[width=0.45\textwidth]{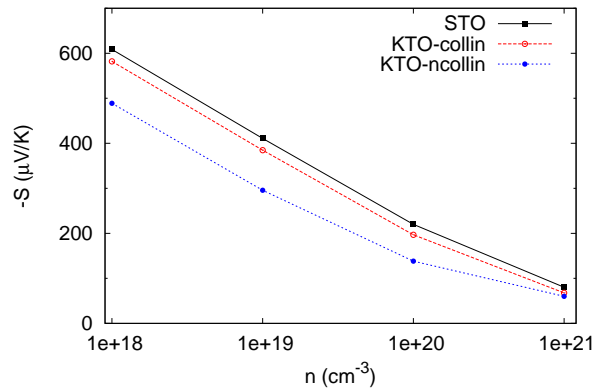}
\caption{\label{fig:Sall} Calculated Seebeck coefficients of STO and KTO at 300K. For KTO,
both collinear and noncollinear calculations are displayed. }
\end{figure}

Fig.~\ref{fig:Sall} shows the calculated Seebeck coefficients as a function of electron concentration at 300 K.
The noncollinear calculation for KTO displays the lowest Seebeck (absolute value) coefficients across the range
of electron densities we have considered. STO on the other hand displays the highest Seebeck coefficients, with
collinear KTO displaying slightly lower values. This behavior can be understood through the dependence of the Seebeck coefficient on
E$_F$. The Seebeck coefficient is inversely proportional to E$_F$~\cite{mahan,seebeck-sto,seebeck-stokto} (for a parabolic band approximation in a 
metal, it is inversely proportional to E$_F$). For a given electron concentration $n$, E$_F$ is lowest for STO, 
since its conduction band width is smaller than KTO, which in turn leads to a higher Seebeck coefficient $|S|$.
The difference between collinear and noncollinear calculations in KTO mainly results from the effective number
of conduction bands. For the noncollinear calculation, the split-up band is higher in energy, leading to an effective
2-conduction-band system (see Fig.~\ref{fig:bands}(b)). For a given $n$, E$_F$ increases when the number of conduction bands decrease; when there are 
less bands which are close in energy, electrons occupy higher lying bands. Thus, $|S|$ in the noncollinear
case is lower than that of the collinear one, due to the reduction of the number of conduction bands~\cite{seebeck-stokto}.

Compared to experimental results, our calculations slightly underestimate the Seebeck coefficients.
For STO, experiments~\cite{seebeck-sto} yield $147\, < |S| < 380$ $\mu V/K$ in the range $10^{21} > n > 8.8 \times 10^{19}$ cm$^3$,
while our calculations yield $80\, < |S| < 220$ $\mu V/K$ in the range  $10^{21} > n >  10^{20}$ cm$^3$.
Similarly for KTO, experiments~\cite{kto-S} yield $200\, < |S| < 460$ $\mu V/K$ in the range
$5.4 \times 10^{18} > n > 1.4 \times 10^{19}$ cm$^3$, while
our calculations yield $138\, < |S| < 489$ $\mu V/K$ in the range $10^{20} > n > 10^{18}$ cm$^3$.
The underestimation of the Seebeck coefficient is most probably related to the overestimation of the 
conduction band-widths in LDA, thus leading to higher E$_F$ which reduces $|S|$~\cite{seebeck-stokto}. In addition, there may be
second order contributions to the Seebeck coefficient from coupling to phonons (i.e. phonon drag effect) missing in our calculations,
although these are most significant at lower temperatures~\cite{sto-ph-drag,stemmer-new}.

The changes in the electron-phonon scattering rates do not lead to a significant change in the calculated
Seebeck coefficients, as we will discuss in the next section. This is in contrast with the calculations
of mobility presented in the previous subsection, which display a strong dependence on the 
values of the scattering rates. The Seebeck coefficient is a ratio
of two transport integrals (see Eqns. (\ref{eqn:tr})). This leads to an exact cancellation 
of the scattering rate when it is taken as a constant. For the case of the scattering rate which is
not constant, this cancellation is not valid. However, our calculations show that the 
difference between the calculations taking into account the full scattering rate $\tau_{n {\bf k}}^{-1}$
and a constant rate is very small for case of room temperature Seebeck coefficients. 

\section{\label{sec:tauk}Constant $\tau$ vs. $\tau_{n {\bf k}}$}

One of the most common approximations in transport calculations is to assume that the scattering rate 
appearing in Eqns.(\ref{eqn:int}) is constant~\cite{boltztrap,boltztrap}. 
For transport coefficients which are ratios of two transport integrals (such as the Seebeck coefficient) 
the dependence on the constant scattering rate disappears.
However, for transport coefficients such as mobility,
the scattering rate is usually fitted to an experimental value. Here, we demonstrate that while a constant scattering rate
provides reasonable Seebeck coefficients, mobilities are not predicted accurately across a range of 
electron densities at 300 K.  

\begin{figure}[!ht]
\includegraphics[width=0.4\textwidth]{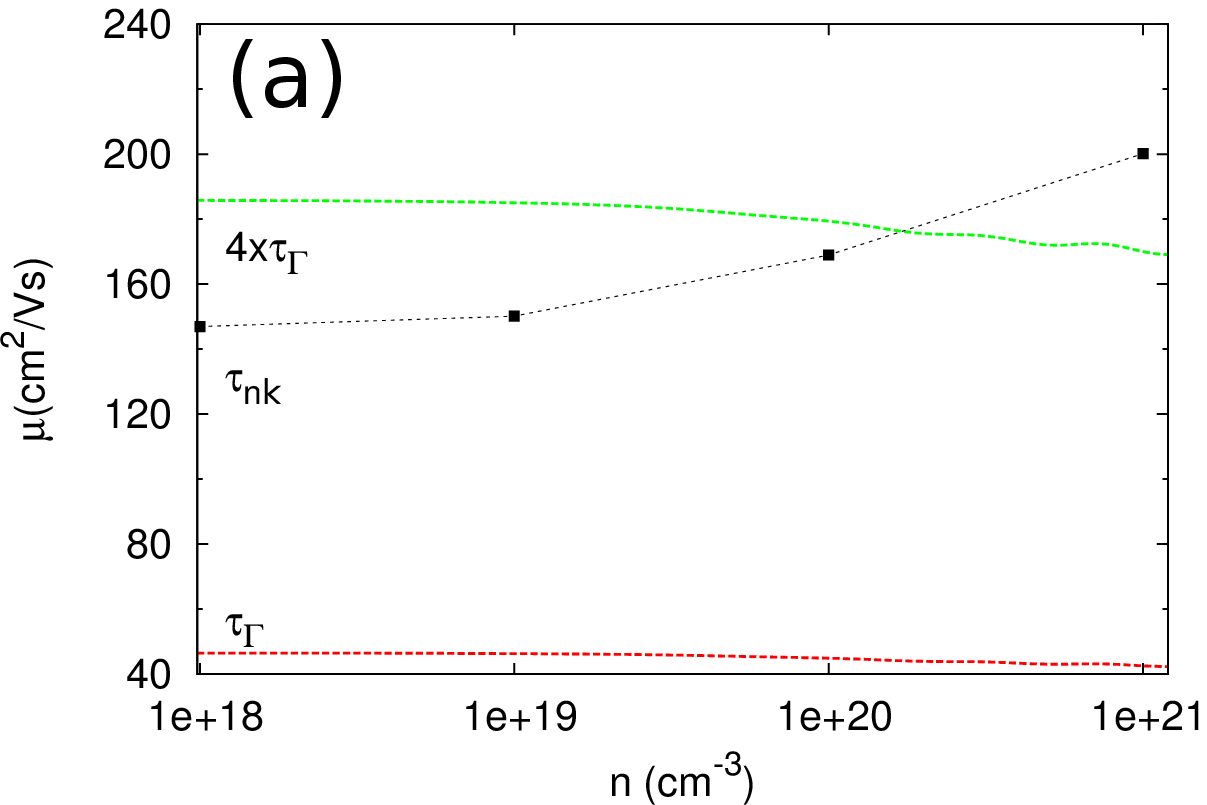}
\includegraphics[width=0.4\textwidth]{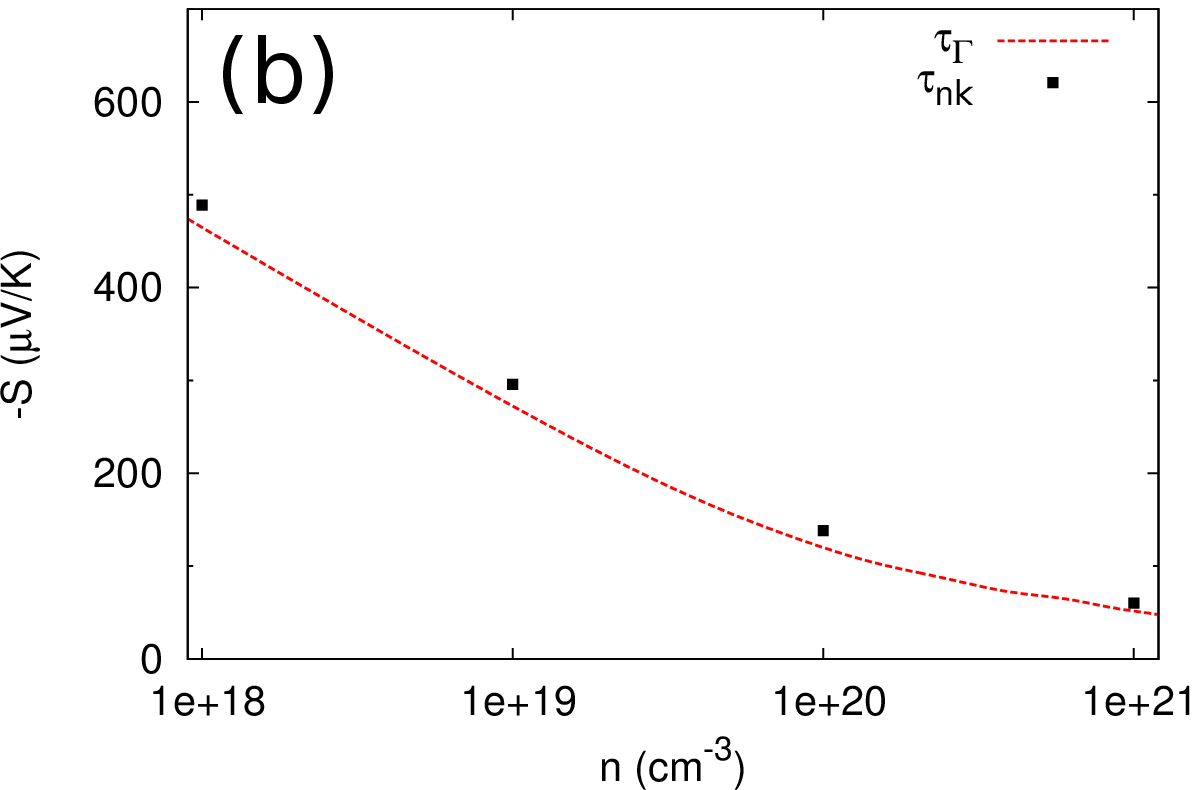}
\caption{\label{fig:MuSKTO} Calculated mobilities (a) and Seebeck coefficients (b) of KTO at 300K as a 
function of $n$ for a constant scattering time $\tau_{\Gamma}$
and the full scattering time $\tau_{n {\bf k}}$. For comparison, mobilities calculated with $4 \times \tau_{\Gamma}$ 
are also shown. }
\end{figure}
In order to compare constant scattering rate with full $k$-dependent scattering rate calculations
presented in the previous section, we choose the constant scattering rate to be the band averaged 
value at the zone center ($\tau_{\Gamma}^{-1}$). The resulting mobilities using a constant scattering rate 
in comparison with the $k$-dependent scattering rate are shown in Fig.(\ref{fig:MuSKTO} a) for 
the noncollinear spin calculation at 300 K (for collinear KTO and STO, similar 
results are obtained, which are not shown here). Also shown in Fig.(\ref{fig:MuSKTO} a) are the
mobilities calculated with constant $0.25\, \tau_{\Gamma}^{-1}$ for comparison. As can be seen,
calculations with constant scattering rate have various problems: 
1)  The average value of the rate at the zone center underestimates the mobility and a larger value 
should be used. 
2) While it is possible to obtain the mobility that results from $k$-dependent scattering rate using just a 
constant rate, this constant number has to be chosen for each given electron density $n$. 
3) The dependence of mobility on $n$ is incorrectly predicted when a constant rate is 
used, since it results in decreasing mobility as a function of $n$, opposite to what 
$k$-dependent scattering rate yields. As a result, a constant rate not only leads to 
a loss of predictability for mobilities (due to not being able to fix a single value for the rate), 
it also leads to incorrect qualitative behavior. 

On the other hand, the Seebeck coefficients computed with a constant rate (which cancels out)
are very close to ones that are computed with full $k$-dependent scattering rate as 
can be seen in Fig.(\ref{fig:MuSKTO} b). This is rather a nontrivial result, since the two transport integrals
in Eqns.(\ref{eqn:int}) are peaked at different regions in the Brillouin zone, where $\tau_{n {\bf k}}$ 
attains different values. In other words, even if we have approximated the integrals $I^{(n)}_{\alpha \beta}$
(n=1,2) with a constant value where the integrands are peaked, the 
scattering times appearing in the numerator and denominator of $S$ will be different and there will be
no cancellation. It is therefore interesting that a constant scattering rate works remarkably well
for predicting room-temperature Seebeck coefficients (assuming all the scattering being due to 
LO phonons) for KTO (and STO). One would expect a similar behavior in other perovskite oxides as well,
thus justifying the use of constant scattering rates for the study of thermoelectric properties.

\section{\label{sec:disc} Conclusion}

In this work, we have studied room-temperature transport properties of the cubic perovskites
KTO and STO. The calculations for the transport properties were based on 
Boltzmann transport theory with relaxation time approximation. The relaxation times were
calculated using an analytical model for scattering of electrons and LO phonons. Our results
have shown that the superior mobility of KTO with respect to STO results from two main reasons:
1) The larger band-width of Ta 5$d$-derived conduction bands in KTO than the Ti 3$d$ derived conduction
bands in STO;
2) Strong spin-orbit coupling in KTO leading to an effective 2-conduction-band system,
compared to the 3-conduction-band system in STO. Despite its larger mobility, KTO has lower Seebeck coefficients, 
due to the fact that the effective number of conduction bands are smaller than in STO. 
We have also compared calculations with constant scattering rate with calculations using
the full $k$-dependent scattering rate. While using a constant scattering rate is a widely
used approximation in the literature, our results have shown that it results in loss of predictability
for calculations of mobility. Instead, for calculations of the Seebeck coefficient, a constant 
scattering rate gives almost the same results with full $k$-dependent scattering rate, 
justifying its use for thermal transport calculations in the literature. 
The computational approach we have presented is rather efficient (compared to those
based on full electron-phonon coupling matrices obtained from DFPT), and results in good agreement
with experiments in predicting the relative transport coefficients of KTO and STO. Using
our approach  to calculate transport coefficients on a large set of perovskite oxides can 
be a basis to search materials with targeted properties for device applications.

\section{Acknowledgements}

We acknowledge support from the Center for Scientific Computing from the CNSI, MRL: an NSF MRSEC (DMR-1121053) and NSF CNS-0960316.
This work used the Extreme Science and Engineering Discovery Environment (XSEDE), which is supported by National Science Foundation grant number ACI-1053575. 
We also thank D. M. Eagles for very useful discussions.

\bibliography{hj}

\end{document}